\documentclass{article}
\usepackage{ijcai25}

\usepackage{times}
\usepackage{soul}
\usepackage{url}
\usepackage[hidelinks]{hyperref}
\usepackage[utf8]{inputenc}
\usepackage[small]{caption}
\usepackage{graphicx}
\usepackage{amsmath}
\usepackage{amsthm}
\usepackage{booktabs}
\usepackage{algorithm}
\usepackage{algorithmic}
\usepackage[switch]{lineno}
\usepackage{utfsym}
\usepackage{multirow} 
\usepackage{graphicx}
\usepackage{array, tabularx, booktabs}
\usepackage{adjustbox}
\usepackage{caption}

\usepackage{multirow} 
\usepackage{amsmath,bm}
\usepackage{soul}
\usepackage{colortbl} 
\setul{0.8pt}{0.8pt}

\definecolor{ourcolor}{rgb}{0.98,0.141,0.183}
\setulcolor{ourcolor}
\sethlcolor{olive!30}
\usepackage{array}    
\usepackage{anyfontsize}
\usepackage{times}  
\usepackage{helvet}  
\usepackage{courier}  
\usepackage{pifont}
\usepackage{graphicx}
\usepackage{array}
\usepackage{tabularx}
\usepackage{caption} 
\usepackage{algorithm}
\usepackage{algorithmic}
\usepackage{booktabs}
\usepackage{threeparttable}
\usepackage{amssymb}
\usepackage[utf8]{inputenc}
\usepackage{array}
\usepackage{rotating} 
\usepackage{afterpage}
\usepackage{subcaption}
 \usepackage{booktabs}
 \usepackage{multirow}

\newcommand{\teai}{\texttt{TEAI} }
\newcommand{\teaiv}{\texttt{TEAI}}
\newcommand{\trai}{\texttt{TRAI} }
\newcommand{\traiv}{\texttt{TRAI}}
\usepackage{newfloat}
\usepackage{listings}
\usepackage{svg}  
\DeclareCaptionStyle{ruled}{labelfont=normalfont,labelsep=colon,strut=off} 
\floatstyle{ruled}
\newfloat{listing}{tb}{lst}{}
\floatname{listing}{Listing}

\newcommand{\rating}[1]{
  \ifcase#1 \ding{108}\ding{109}\ding{109}\ding{109}\ding{109} \or
  \ding{108}\ding{108}\ding{109}\ding{109}\ding{109} \or
  \ding{108}\ding{108}\ding{108}\ding{109}\ding{109} \or
  \ding{108}\ding{108}\ding{108}\ding{108}\ding{109} \or
  \ding{108}\ding{108}\ding{108}\ding{108}\ding{108}
  \fi
}

\usepackage{fancyhdr}
\fancypagestyle{firstpage}{
    \fancyhf{} 
    \fancyhead[C]{\small \textbf{Published version:} \\ 
    This paper has been accepted for publication at IJCAI 2025. \\ 
    The final version is available at \url{https://www.ijcai.org/proceedings/2025/1066.pdf}}
}

\title{Towards the Terminator Economy: Assessing Job Exposure to AI Through LLMs}

\author{
Emilio Colombo$^{1,3}$
\and
Fabio Mercorio$^{2,3}$\and
Mario Mezzanzanica$^{2,3}$\And
Antonio Serino$^4$\\
\affiliations
$^1$Dept of International Economics, Institutions and Development, Catholic University of Milan\\
$^2$Dept of Statistics and Quantitative Methods, University of Milano-Bicocca, Italy\\
$^3$CRISP Research Centre, University of Milano-Bicocca, Italy\\
$^4$Dept of Economics, Management and Statistics, University of Milano-Bicocca\\
\emails
emilio.colombo@unicatt.it,
\{fabio.mercorio, mario.mezzanzanica, antonio.serino\}@unimib.it,
}
\begin{document}

\maketitle
\thispagestyle{firstpage} 

\begin{abstract}
AI and related technologies are reshaping jobs and tasks, either by automating or augmenting human skills in the workplace. Many researchers have been working on estimating if and to what extent jobs and tasks are exposed to the risk of being automatized by AI-related technologies. Our work tackles this issue through a data-driven approach by:  
(i) developing a reproducible framework that uses cutting-edge open-source large language models to assess the current capabilities of AI and robotics in performing job-related tasks;  
(ii) formalising and computing a measure of AI exposure by occupation, the \teai (\texttt{T}ask \texttt{E}xposure to \texttt{AI}) index, and a measure of \texttt{T}ask \texttt{R}eplacement by AI, the \trai index, both validated through a Human user evaluation and compared with the state-of-the-art. 

Our results show that the \teai index is positively correlated with cognitive, problem-solving, and management skills, while it is negatively correlated with social skills. Results also suggest about one-third of U.S. employment is highly exposed to AI, primarily in high-skill jobs requiring a graduate or postgraduate level of education. We also find that AI exposure is positively associated with employment and wage growth in 2003--2023, suggesting that AI has had an overall positive effect on productivity.  
Considering specifically the \trai\ index, we find that even in high-skill occupations, AI exhibits high variability in task substitution, suggesting that AI and humans complement each other within the same occupation, while the allocation of tasks within occupations is likely to change.

All results, models, and code are freely available online to allow the community to reproduce our results, compare outcomes, using our work as a benchmark to monitor AI’s progress over time.
\end{abstract}

\section{Introduction}
Artificial Intelligence (AI) provides strong arguments for being considered a general-purpose technology due to its broad applicability, potential productivity gains and possibility to drive further innovation \cite{Crafts-21,Brynjolfsson-Rock-Syverson-19,deming2025technological,WEF2025}.
However, these characteristics of AI create a relevant measurement problem, as it is extremely difficult to identify all the channels through which it affects the economy. Meanwhile, in late 2018, Large Language Models (LLMs) were introduced as powerful computational tools designed to understand and generate human-like text by leveraging vast amounts of data, see \cite{min2023recent}. Due to their performance and adaptability as off-the-shelf models, they have become central to numerous systems capable of processing and generating text, driving applications in natural language understanding, translation, and content creation across domains.

\paragraph{Contribution.} This paper develops a methodology for assessing AI exposure using LLMs, with a granular approach that analyses exposure for each job task within each job occupation from two different perspectives:

\emph{(i) From a methodological point of view}, we design and implement a reproducible framework 
    to estimate to what extent existing AI and robotics technologies can perform job-related tasks relying on LLMs. Instead of assessing AI exposure through external benchmarks such as expert judgment or AI patents and innovations data, we construct an internal assessment using LLM's own evaluation, using O*NET\footnote{O*NET is the official US comprehensive taxonomy of about 1K occupations and 19K+ job-related tasks} as reference occupation/task taxonomy. The result is a framework that enables individuals and organizations to develop policies for integrating AI into careers and the workforce.

\emph{(ii) From an economic perspective}, we develop a measure to estimate the AI exposure and replacement index by tasks for all the occupations of the US taxonomy. We conduct an in-depth analysis of the impact of AI across occupations in the US labor market, examining job characteristics such as skill levels and educational requirements. We investigate the relationship between AI exposure and labor market outcomes like employment and wage growth over 2019–2023. This combined analysis allows us to estimate the impact of AI and technology on different occupations and economic indicators.

\paragraph{Reproducibility and Demo.} Codes and results are available on GitHub\footnote{\url{https://github.com/Crisp-Unimib/Terminator-Economy}} while an interactive online demo is available at \url{http://terminatoreconomy.com/}

 \section{Related Works on AI and Jobs\label{sec:literature}}

Since the seminal paper by~\citeauthor{Autor-Levy_Murnane-2003}, the task approach has  been applied to analyze the effect of technology and trade (offshoring) \cite{Acemoglu-Autor-11}, to the long run effect of technology \cite{Consoli-Marin-Rentocchini-Vona-2023}, and to skill-task interaction \cite{Colombo-Mercorio-Mezzanzanica-2019}.

More recently, this approach has been used to measure occupational exposure to computers and robots. In a seminal paper  \cite{Frey-Osborne-17} estimated that up to 47\% of jobs in the US are at risk of automation.\footnote{See also \cite{Nedelkoska-Quintini-18} for a similar approach.} Subsequently, other attempts focused on developing measures of exposure to machine learning and robotics \cite{Brynjolfsson-Mitchell-17,Acemoglu-Restepo-2020} and to AI \cite{Felten-Raj-Seamans-21,Webb-23,Eloundou-Manning-Mishkin-Rock-23,Pizzinelli-Panton-Tavares-Cazzaniga-Li-23}.

A common feature of those is the attempt to quantify AI exposure through an \textit{external} benchmark, which may be expert judgment or data analysis on patents and innovations. 

In contrast, our approach is based on an \textit{internal} assessment, where LLM systems are asked to assess the suitability of tasks for AI. This approach has two main advantages. First, it is fully transparent, with results and outcomes being fully disclosed. Second, the approach is fully reproducible. This means that whenever new generations of LLMs are available, they can be used in our approach to measure the change in task exposure that they imply.

In February 2025, Anthropic published a study\footnote{\url{https://www.anthropic.com/news/the-anthropic-economic-index}}  to assess AI’s impact on jobs and tasks using O*NET, mapping millions of Claude conversations to O*NET tasks. By contrast, our approach (i) maps \textit{every} O*NET occupation and task, (ii) employs a panel of three open LLMs to ensure reproducibility and minimize hallucinations, and (iii) allows the community to estimate the AI impact on other third-party taxonomies.

\section{Building the AI Exposure and Substitutability Indexes \label{sec:ai-index}}
\begin{figure}[ht]
    \centering
\includegraphics[width=0.45\textwidth]{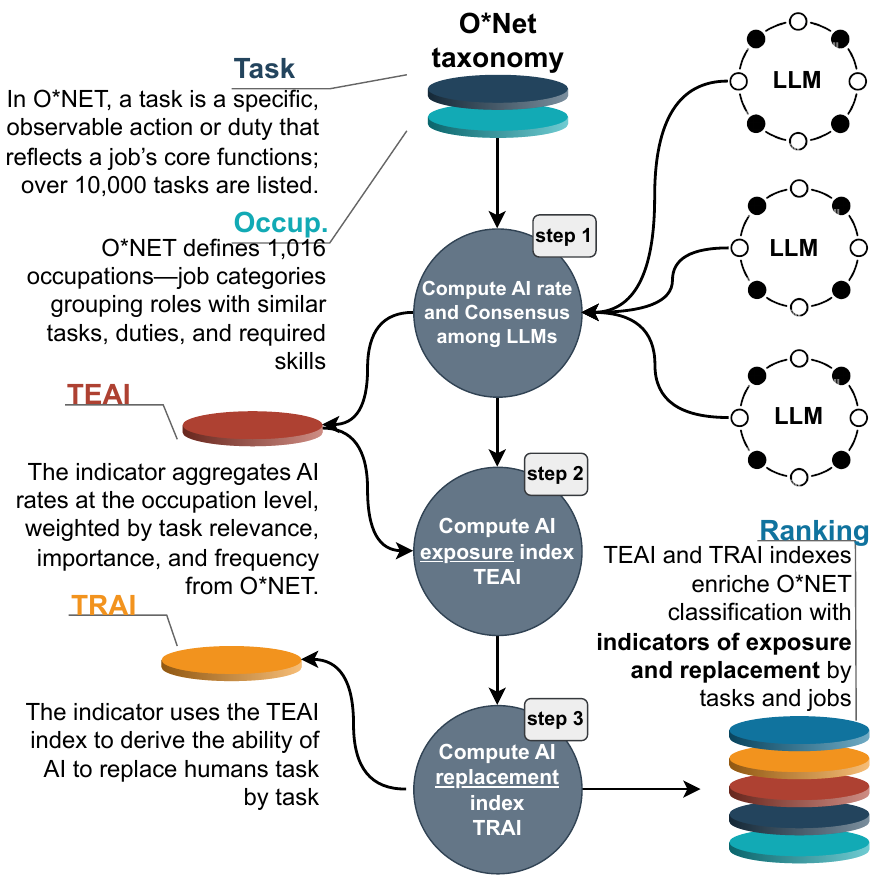}
    \caption{Graphical overview of the framework}
    \label{fig:framework}
\end{figure}

Our method can be summarised as in Fig.~\ref{fig:framework}.


\subsection{Step~1: Compute the AI Rate} 
\label{sec_teai}
To mitigate the risk of being driven by the well-known problem of LLMs, known as "hallucinations" \cite{ji2023survey}, we design a framework involving three different LLMs that aim to identify and limit the false information generated, thereby creating a consensus system among them.

\paragraph{Model choice.}
To ensure the reproducibility of this work, we use three of the best-performing open-source models (at the time of the experimental phase), according to performance benchmarks, available in the open LLM ranking.\footnote{\url{https://huggingface.co/spaces/HuggingFaceH4/open_llm_leaderboard}}
The three selected models are \textit{Mistral 7B Instruct v 0.2} \cite{jiang2023mistral}, \textit{openchat 3.5 0106} \cite{wang2023openchat} and \textit{orca mini v3 7b} \cite{mukherjee2023orca}. 
\paragraph{Prompt design.} The starting point is the O$^*$NET taxonomy which identifies 19281 tasks for 923 SOC occupations.\footnote{We use the O$^*$NET 28.2 version released in February 2024.}
We formulate a five-shot prompt using the few-shot learning approach \cite{brown2020language}.
We use each individual task description assigned to an occupation to ask the models how well, on a scale of 1 to 5,\footnote{How well an AI system, which can be an LLM, Image Processing System or a Robot, could perform in the task on a scale of 1 to 5 where 1 stands for \textit{poor} and 5 stands for \textit{excellent}?} the combination of different AI technologies could perform the input task and provide a discursive motivation for the evaluation. 
As AI technologies, we consider i) LLMs for textual data understanding, ii) Image Processing Systems for elaboration and decision-making based on visual data analysis, and iii) Robotic systems for physical execution.
At the end of the prompt, the model is provided with five examples of the task to obtain more contextual and accurate results.

\paragraph{Model results.}
We iterate this process for each task provided by O$^*$NET and for each model chosen, resulting in three scores and natural language motivations provided by each model. Table \ref{tab:table1} provides an example of the results after this stage for a selection of occupations and tasks.



\begin{table}[h!]
\tiny
\begin{tabular}{|l|c|c|c|c|c|}
\hline
\multicolumn{1}{|l|}{\textbf{O*NET Task}} &
  \multicolumn{1}{l|}{\textbf{Mistral}} &
  \multicolumn{1}{l|}{\textbf{\begin{tabular}[c]{@{}l@{}}Orca\\ mini\end{tabular}}} &
  \multicolumn{1}{l|}{\textbf{\begin{tabular}[c]{@{}l@{}}Open\\ Chat\end{tabular}}} &
  \multicolumn{1}{c|}{\textbf{\begin{tabular}[c]{@{}c@{}}Motivation \\ Mistral\end{tabular}}} &
  \multicolumn{1}{l|}{...} \\ \hline
\begin{tabular}[c]{@{}l@{}}Direct or \\coordinate \\the\\ supportive \\ services \\department \\of\\ a business \\ agency, or \\organization.\end{tabular} &
  4 &
  4 &
  3 &
  \begin{tabular}[c]{@{}c@{}}Robotics can automate\\ repetitive tasks such as \\\hl{data entry}, \hl{document scanning}, \\and \hl{mail sorting}. Image\\ Processing Systems can \\\hl{analyze documents} and \hl{emails} \\ for information extraction and\\ categorization. LLMs can handle\\ complex tasks like \hl{scheduling}, \\\hl{coordinating resources}, and\\ \hl{communicating} with team \\members. Combining these \\AI technologies can significantly \\improve the efficiency and \\accuracy of administrative \\services, allowing managers to\\ focus on more strategic tasks.\end{tabular} &
  ... \\ \hline
\begin{tabular}[c]{@{}l@{}}Conduct \\classes to \\teach\\ procedures \\to staff.\end{tabular} &
  2 &
  3 &
  2 &
  \begin{tabular}[c]{@{}c@{}} LLMs could potentially\\ \hl{create instructional materials},\\ but they \hl{lack} the \hl{ability} to\\ \hl{engage} and \hl{assess} staff \\performance during classes. \\Image Processing Systems\\have no role in this task.\\ Robotics \hl{could theoretically}\\ \hl{conduct classes} through\\ pre-recorded videos, but\\ the human interaction and \\customization required for \\effective training make this a \\poor solution.\end{tabular} &
  ... \\ \hline
\end{tabular}%
\caption{Extraction of two tasks for \href{https://www.onetonline.org/link/summary/11-3012.00}{\textit{Administrative Services Managers}} occupation O*NET, enriched as discussed in Step1 to derive to what extent existing AI and robotics technologies can perform job-related tasks according to Mistral, Orca-mini, and Open-Chat LLMs. The rating scale expressed by the LLMs corresponds to the following categorical labels: 1: Poor, 2: Fair, 3: Average, 4: Good, 5: Excellent. \hl{Key activities are highlighted for a better reading}}\label{tab:table1}
\end{table}

\paragraph{Evaluating Consensus among LLMs.}
As mentioned above, the decision to use three different models was made to limit hallucinations. 
In Section~\ref{consensus chap}, we describe the analysis conducted on the output of the three models and how we obtain a single indicator for each task, defined as \textsc{te}, a metric ranging from 1 to 5 that measures the extent to which AI can perform each specific task.

\subsection{Step~2: Compute the AI Exposure}

To compute occupation exposure to AI, we follow \cite{Felten-Raj-Seamans-21} and aggregate the \textsc{te} scores at the occupation level by weighting them by task relevance ($R$), importance ($I$), and frequency ($F$) as measured by O$^*$NET.\footnote{Weights capture different aspects of the tasks. Importance: indicates the degree to which a particular descriptor is important for the occupation. 
Relevance refers to the proportion of incumbents who rated the task as relevant to their job.
Frequency refers to the frequency of each task within the occupation, from annual to hourly. 
} For each task $j$ and occupation $i$ our AI exposure score is computed as follows:

\begin{equation}
    TEAI_i = \frac{\sum^n_{j=1} TE_{ij} \cdot R_{ij} \cdot I_{ij} \cdot F_{ij} }{\sum^n_{j=1} \cdot R_{ij} \cdot I_{ij} \cdot F_{ij}}
    \label{eq:teai}
\end{equation}

where $TE_{ij}$ identifies the metric developed in step 1 at the task level, $n$ defines the number of tasks within each occupation. Each weight is scaled by its maximum to obtain equal weights. The O$^*$NET model uses different scales for Relevance (scale 1-100), Importance (scale 1-5), and Frequency (scale 1-7). We normalize the indices to ensure equal scale across weights. 
Finally, the score is normalized to ensure comparability with other similar scores.

\subsection{Step~3: Compute the AI Sustitutability}
The \teai index measures exposure to AI but does not specifically address the issue of task substitution or replacement. To deal with this, we leverage information contained in the motivations obtained in the first step. 

We synthesize these, generating a consolidated motivation that summarizes the different justifications regarding the AI exposure rate for each occupation-related task. 
To obtain a quantitative measure of AI's involvement in each task, we provide this summarized motivation along with the task title and the associated job occupation title to another LLM,
  prompting it to return the level of AI engagement on a scale from 1 (``no engagement'') to 5 (``complete automation, replacing human execution'') and the motivation regarding the engagement. 
\paragraph{Model choice and Results.} In this step we used a single LLM that was larger than those used in the first step. To ensure the complete reproduction of the results, we chose the \textit{Qwen 72B} model, which, among open-source models, represents a good trade-off between size and performance.
The result is a measure of task substitutability that we aggregated at the occupation level using eq \eqref{eq:teai}, obtaining an index of Task Replacement by AI~(\traiv).  

\section{Human Evaluation of TEAI/TRAI Indexes}
To evaluate the outcomes provided by LLMs, we conducted a user study by recruiting 12 evaluators through Prolific~\cite{palan2018prolific}, an online service offering
participants for research studies. It has been recognized as a cost-effective platform that provides high-quality data for online research, making it suitable for studies where reliability and data integrity are priorities~\cite{douglas2023data}, and used in several computer-science related user evaluations, see, e.g.,~\cite{de2024large,koppol2021interaction}. 
\paragraph{Setting the user study.} We sampled occupations from the O*NET database by SOC first-digit, selecting 20\% of occupations and choosing two tasks for each occupation: one with the highest and one with the lowest TEAI/TRAI indexes. This resulted in a total of 200 occupations and 400 tasks. The sample was further divided into four chunks of 100 tasks each. Each sample was composed of job occupation title, task title, \teai value, \teai natural language motivation, \trai value, and \trai natural language motivations.\footnote{To avoid potential biases in human judgment, the evaluators were not informed that AI systems had generated the indexes and summaries they reviewed} 
Each chunk was assigned to exactly three participants, whom we asked to consider each task for each job occupation and evaluate the \teai and \trai, with a nominal value\footnote{I disagree with both; I agree only with TEAI; I agree only with TRAI; I agree with both} indicating their degree of agreement with the indices. 
We aggregate the results, obtaining agreement performance for \teai and \traiv. It is important to highlight that these results are not intended to represent true accuracy, but a consistency with what our methodology has estimated. Consensus among the participants was also calculated using the equation~\ref{eq:consensus} to assess the extent to which participants agreed on the assignment of evaluation values.
Results are shown in Tab.~\ref{tab:accuracy_agreement}.

\paragraph{Results.} Averagely, evaluators tend to agree with (i) \teai index 75\% of the time (48\%+27\%); (ii) 71\% of the times with \trai (48\%+23\%), whilst (iii) only 2\% of the times they disagree with the indicators. However, the reader may notice that the level of consensus among reviewer is lower than expected. This is largely because the topic at hand — the impact of AI on tasks and professions — is inherently subjective and influenced by human variability, convictions, and expectations, making evaluations highly personal and, at times, open to interpretation.
Nonetheless, the results show that our indices are meaningful as they effectively capture the current capabilities of state-of-the-art AI technologies in supporting and replacing tasks within various jobs. As a result, \teaiv, \traiv, and the natural language explanations provide a valuable framework for individuals and organizations to develop policies that integrate AI into careers and the workforce.
\setlength{\tabcolsep}{1.7pt}  

\begin{table}[ht]
    \centering
    \renewcommand{\arraystretch}{1.2}
    \resizebox{.48\textwidth}{!}{  
    \begin{tabular}{@{}l|cccc>{\columncolor[gray]{0.9}}c|cccc>{\columncolor[gray]{0.9}}c@{}}
        \cline{2-11}
        & \multicolumn{5}{c|}{\textbf{Coherence by Chuck ( \%)}} & \multicolumn{5}{c}{\textbf{Consensus by Chuck}} \\ 
        \cline{2-11}
        & C1 & C2 & C3 & C4 & \textbf{Mean} & 
          C1 & C2 & C3 & C4 & \textbf{Mean} \\ 
        \midrule
        \textbf{Agree with both} & 35.84 & 53.76 & 32 & 65.94 & 48  & 
                              .69 & .72 & .54 & .72 & .68 \\ 
        \textbf{Agree more with \teai} & 27.96 & 19 & 45.48 & 17.03 & 27  & 
                              .66 & .57 & .67 & .52 & .61 \\ 
        \textbf{Agree more with \trai} & 35.13 & 26.88 & 20.35 & 11.23 & 23 & 
                              .75 & .76 & .70 & .64 & .71 \\ 
        \textbf{Disagree with both}   &  1.08 &  0.36 &  2.17 &  5.89 &  2  & 
                              .50  & .27 & .41 & .28 & .36  \\ 
        \bottomrule
    \end{tabular}
    } 
    \caption{Human evaluation results. Mean of Coherence and Consensus Levels among reviewers by chunk. Each chunk has been assigned to three reviewers. The closer the consensus is to 1, the more the different reviewers agree.}
    \label{tab:accuracy_agreement}
\end{table}


\section{Experimental Results and Outcomes on the US Labor Market \label{sec:results}}
\begin{figure}[htb]
    \centering
    \includegraphics[width=0.9\linewidth]{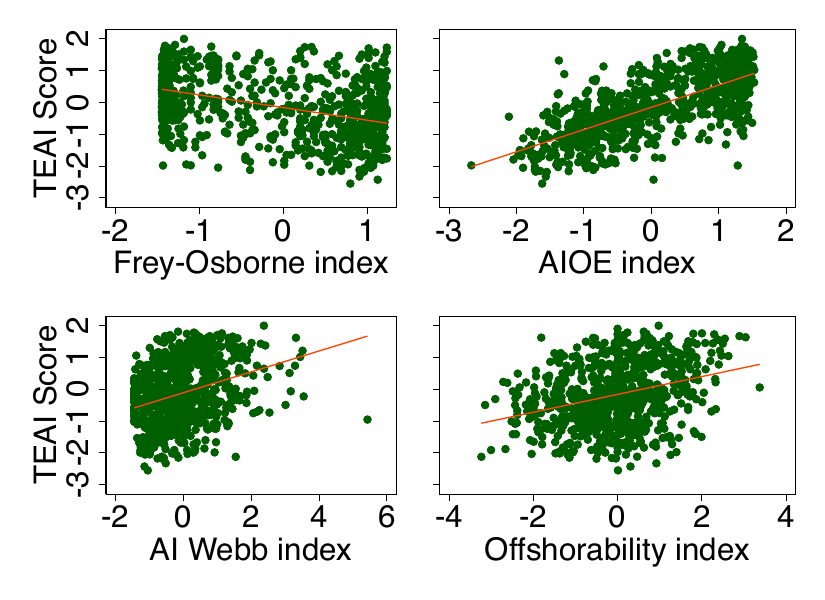}
    \caption{Correlation with existing exposure indexes. Each dot represents a SOC occupation.}
    \label{fig:scatter_te_1a}

\end{figure}

\paragraph{Benchmarking evaluation.} First, we compare our AI index with other existing measures in the literature. Figure \ref{fig:scatter_te_1a} shows the correlation between the \teai index and the well-known measure developed by~\citeauthor{Frey-Osborne-17}, the AI exposure index by~\citeauthor{Felten-Raj-Seamans-21} and by~\citeauthor{Webb-23}, and the offshorability index developed by~\citeauthor{Acemoglu-Autor-11}.  The pairwise correlation is always statistically significant at 5\%. It is higher for the AIOE index, much lower for the AI Webb and the offshorability index, and negative for the Frey-Osborne index. This means our measure is broadly consistent with existing measures but captures different elements of the relationship between AI and the labor market. The negative correlation with the Frey and Osborne index can be explained by the latter being a measure of exposure to robotics and computerisation, and more focused on routine tasks. At the same time, generative AI is more focused on cognitive/non-routine tasks. 

\begin{figure}[htb]
    \centering
    \includegraphics[width=\linewidth]{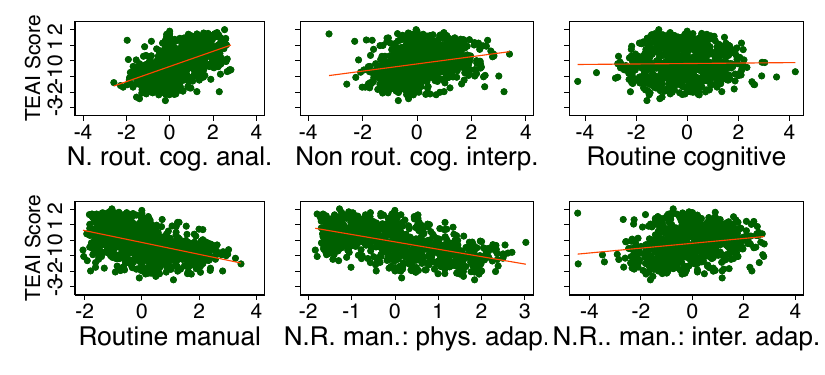}
    \caption{Correlation with different skill intensity measures. }
    \label{fig:scatter_te_2a}
\end{figure}

\subsection{AI and Skills}
Next, we explore the relationship between our \teai index and different skills. Figure \ref{fig:scatter_te_2a} displays scatterplots comparing the \teai index with the intensity of different skill types at occupation level derived from \cite{Acemoglu-Autor-11}. The graph shows the peculiar nature of AI technologies, which are positively correlated with cognitive analytical and interpersonal skills but negatively correlated with routine manual skills and non-routine manual skills that require physical adaptability. Surprisingly, the correlation with cognitive routine skills is only weakly positive, while it is positive for non-routine manual skills requiring interpersonal adaptability. The results of the figure are purely descriptive, therefore we add a more robust analysis by extracting from O$^*$NET the detailed skills associated with each occupation. We group the skills into 4 classes: Cognitive, Social, Problem Solving and Management, and Technical skills. We then develop a skill relevance index for each class at the occupation level by weighting each skill according to its level and importance.\footnote{As provided by O$^*$NET.} The skill relevance index is constructed as follows: 

\begin{equation}
    SR_{ci} = \frac{\sum^m_{z=1} S_{zcj} \cdot L_{zcj} \cdot I_{zcj} }{\sum^m_{z=1} \cdot L_{zcj} \cdot I_{zcj}}
    \label{eq:sr}
\end{equation}

where $z$ denotes the $m$ skills of class $c$ in each occupation $j$; $L$ and $I$ denote, respectively, the level and importance of each skill in each occupation. 

We then estimate the following regression:

\[
TEAI_i = \alpha + \bm{S}^\prime_i \bm{\beta} + \gamma O_i + \epsilon_i
\]

where each observation is a SOC occupation ($i$), $TEAI_i$ is our measure of AI exposure, $\bm{S}$ is a 4x1 column vector of skill relevance for each of the skill classes described above at the occupation level, $\bm{\beta}$ is a 4x1 column vector of coefficients, $O$ defines occupation dummies and $\epsilon_{i}$ the error term. We saturate the model with more detailed dummies up to the fifth digit; therefore, the results are identified within each group variation. Table \ref{tab:reg_te_skill} shows the results. The \teai index is positively correlated with cognitive, problem-solving, and managerial skills, but negatively correlated with social skills, as expected. The relationship with technical skills is very weak and does not survive the inclusion of detailed SOC occupation dummies.  

\begin{table}[!ht]
        \centering
      \begin{threeparttable}
{
\def\sym#1{\ifmmode^{#1}\else\(^{#1}\)\fi}
\tiny
\begin{tabular}{l*{5}{c}}
\toprule
\hline
                    &\multicolumn{1}{c}{(1))}&\multicolumn{1}{c}{(2)}&\multicolumn{1}{c}{(3)}\\
\hline
Cognitive           &      5.6934\sym{***}&      5.3653\sym{***}&      5.3105\sym{***}\\
                    &    (0.8023)         &    (0.7937)         &    (1.1712)         \\
\addlinespace
Social              &     -3.1395\sym{***}&     -3.2464\sym{***}&     -3.0310\sym{**} \\
                   &    (0.7239)         &    (0.7335)         &    (1.1295)         \\
\addlinespace
Management &      3.1883\sym{***}&      3.4810\sym{***}&      2.9411\sym{*}  \\
                     &    (0.8183)         &    (0.8129)         &    (1.3644)         \\
\addlinespace
Technical           &      0.0259         &     -0.0419         &     -0.1486         \\
                    &    (0.6432)         &    (0.6527)         &    (0.8556)         \\
\hline
SOC FE              &          3d         &          4d         &          5d         \\
R2                  &       0.775         &       0.780         &       0.854         \\
N                  &         774         &         771         &         522         \\
\hline
\bottomrule
\end{tabular}
}

      \end{threeparttable}  
\caption{OLS estimates of TEAI index on measures of skill intensity.   \textit{Note:} Each observation consists of an occupation. OLS regression using \teai index as the dependent variable.  The independent variables are skill intensities.  All the regressions include occupation (SOC) fixed effects at 3, 4, and 5 digits. Robust standard errors in parentheses *** p $<$ 0.001, ** p $<$ 0.01, * p $<$ 0.05}     \label{tab:reg_te_skill}
\end{table}
\subsection{AI Employment and Wages}

Finally, we explore the relationship between \teai and labor market outcomes. We begin by analyzing the size and characteristics of workers exposed to AI technologies. First, we divide the distribution of \teai scores into three tertiles representing high, medium, and low AI exposure. We then compute the degree of exposure of the US population using BLS employment data. Finally, we distinguish between occupational groups and by skill groups within each tertile.

\begin{figure}[h!]
    \centering
    \begin{subfigure}[t]{0.4\textwidth}
    \includegraphics[width=\linewidth]{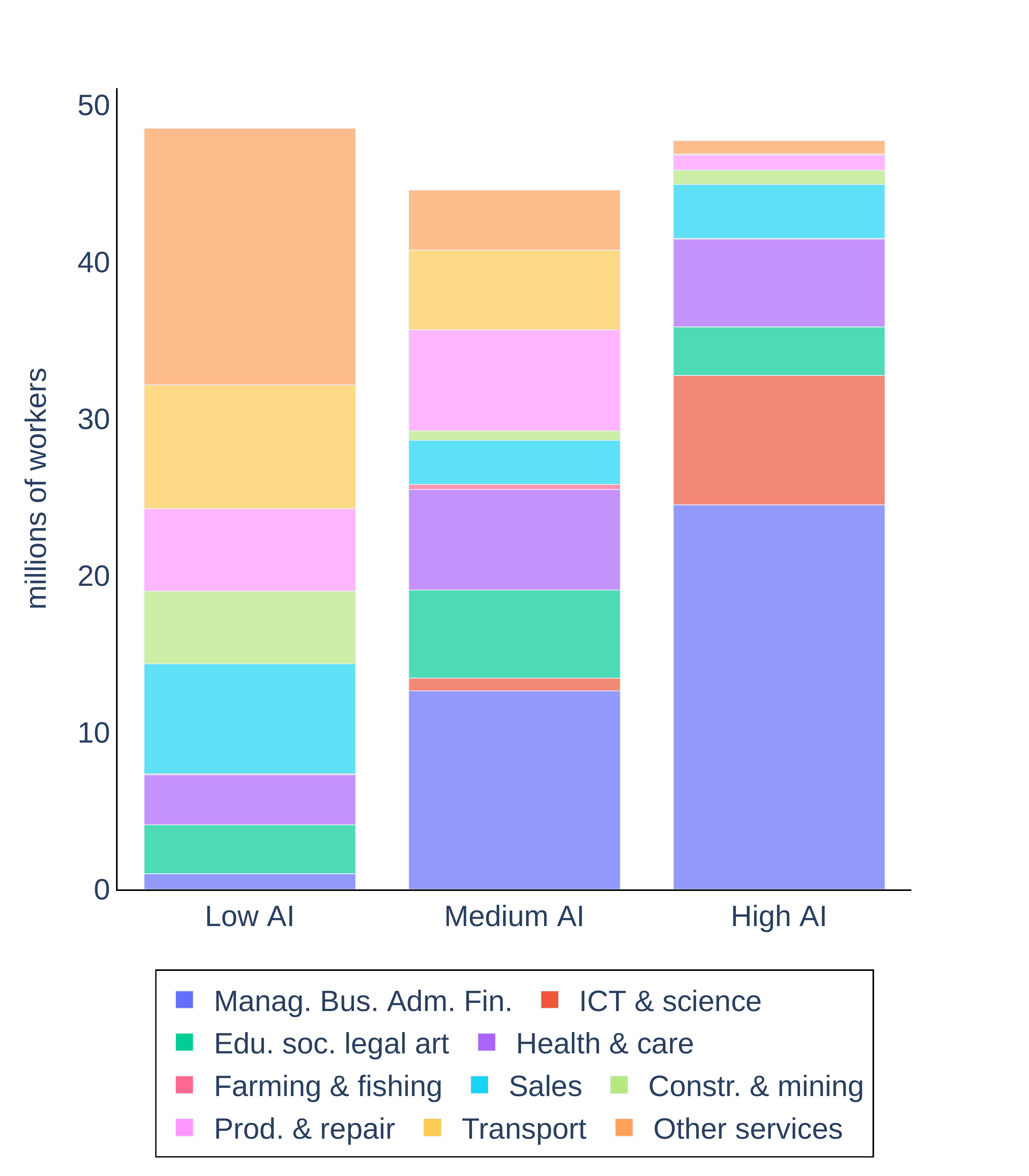}
    \caption{\teai index by SOC group of O*NET}
    \label{fig:dist_te_emp_soc_sec}
    \end{subfigure}%
    \\
    \begin{subfigure}[t]{0.4\textwidth}
    \includegraphics[width=\linewidth]{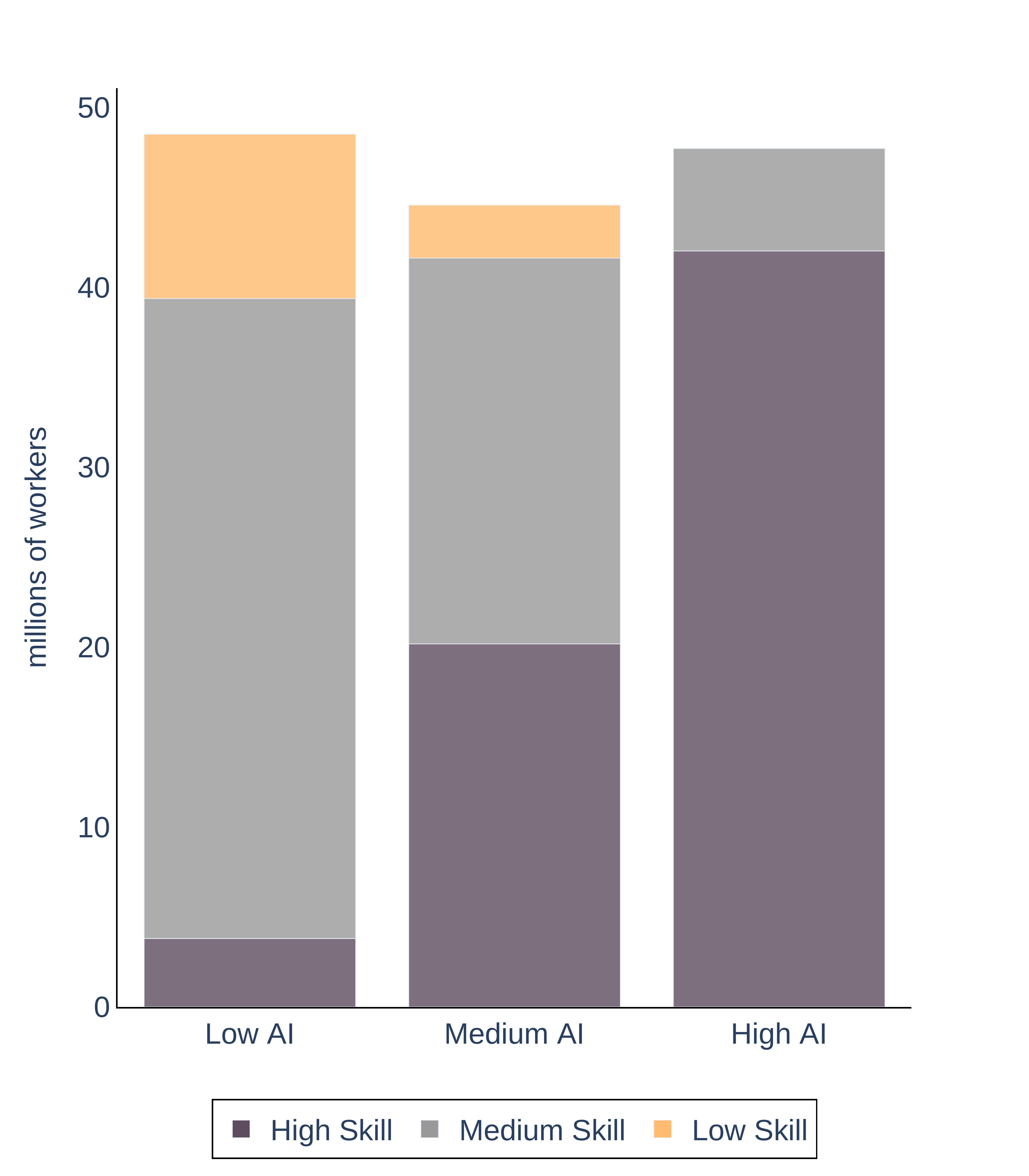}
    \caption{\teai index by skill intensity}
    \label{fig:dist_te_emp_s}   
    \end{subfigure}
    \caption{Figures~\ref{fig:dist_te_emp_soc_sec} and~\ref{fig:dist_te_emp_s} display the distribution of exposure to the \teai\ index across SOC groups and skill intensity levels, respectively, using U.S. BLS employment data (in millions of workers). For both panels, each bar represents a tertile of the \teai\ score distribution.} 
\end{figure}

\begin{figure}[htb]
    \centering

    \begin{subfigure}[t]{0.48\textwidth}
        \includegraphics[width=\linewidth]{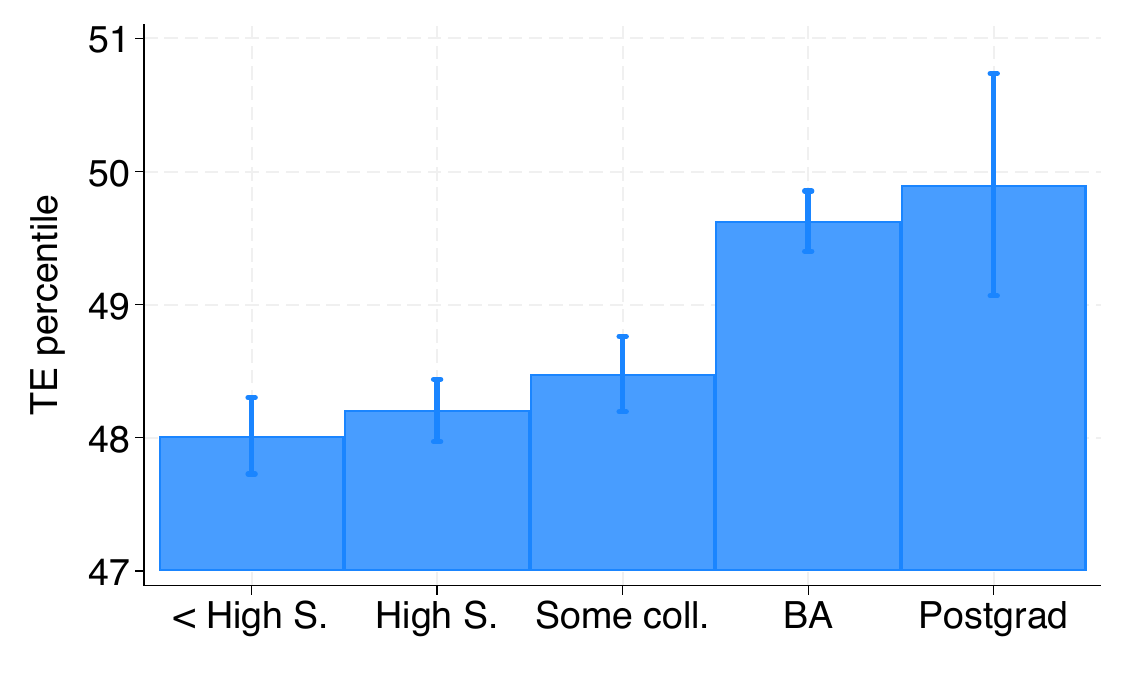}
        \caption{Exposure by education}
        \label{fig:marginsplot_educ}
    \end{subfigure}
    \\
    \begin{subfigure}[t]{0.48\textwidth}
         \includegraphics[width=\linewidth]{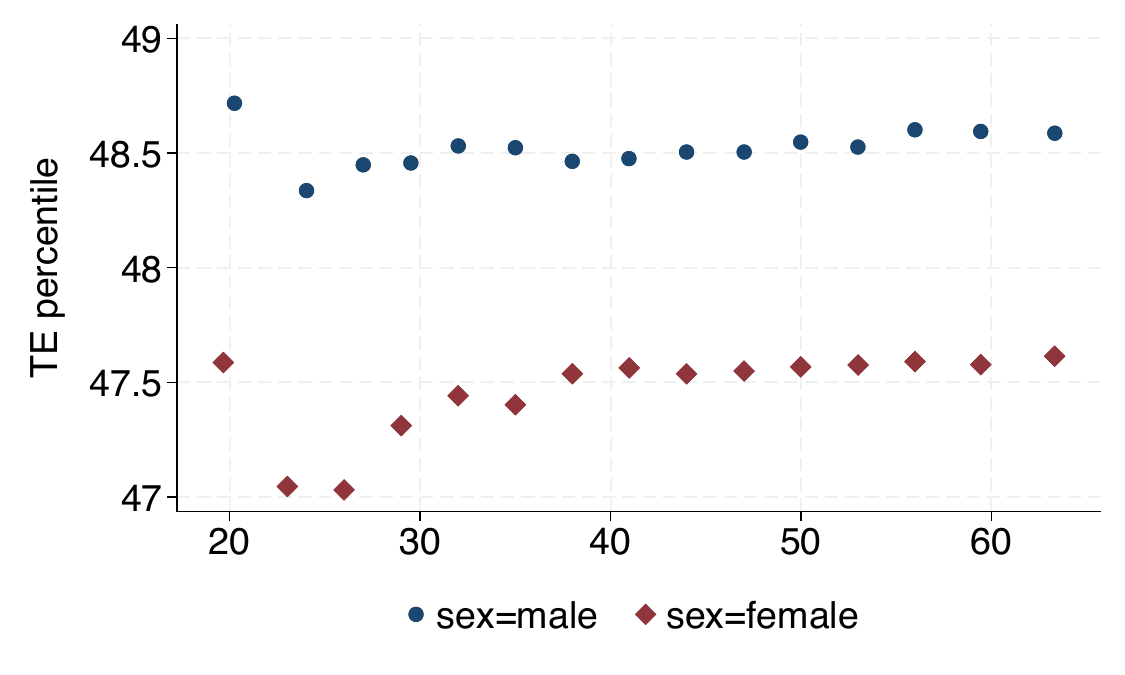}
        \caption{Exposure by age and sex}
        \label{fig:binscatter_age}   
    \end{subfigure}
    \caption{AI exposure by workers' characteristics.} 
    \label{fig:margins_biscatter}

    \caption*{\footnotesize{(a) shows coefficients of regression of education categories on \teai exposure. Controls  include age, sex, occupation(4d), industry(3d), state and year fixed effects. 
    (b) is a binscatter. The x-axis is the average age of workers in an industry-occupation-state observation in the 2022-18 ACS 5-year sample. Biscatter is computed considering education as a covariate. ACS individual weights are used.}}
\end{figure}

\begin{figure}[htb]
    \centering
    \begin{subfigure}[t]{0.48\textwidth}
        \includegraphics[width=\linewidth]{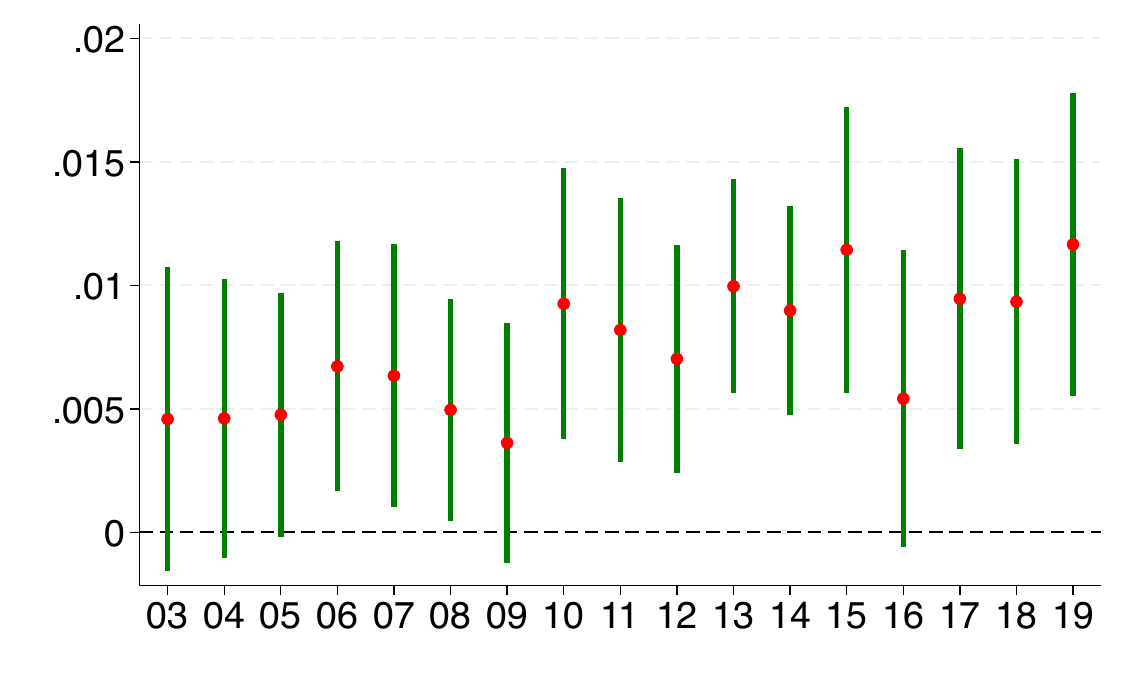}
        \caption{\teai index and employment growth }
        \label{fig:coefplot_demp}
    \end{subfigure}
    \\
    \begin{subfigure}[t]{0.48\textwidth}
        \includegraphics[width=\linewidth]{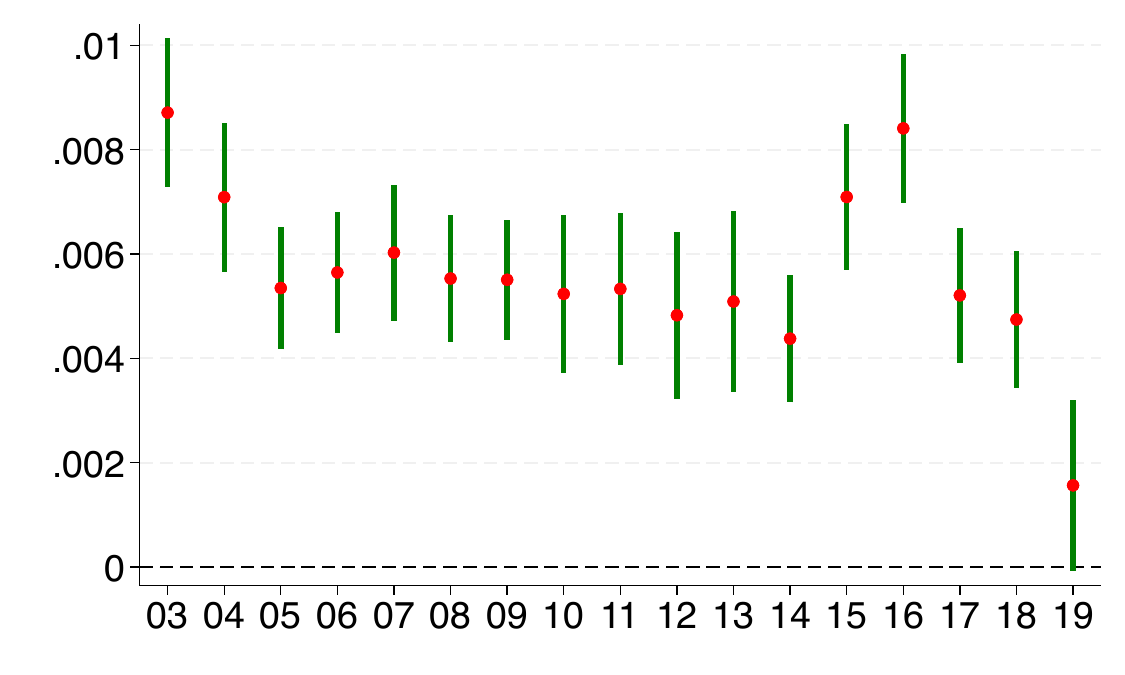}
        \caption{\teai and wage growth}
        \label{fig:coefplot_daw}   
    \end{subfigure}
       \caption{\teai index, employment and wage growth.} 
    \label{fig:rolling_emp_wage}
 
    \caption*{\footnotesize{This figure plots the effect of AI score on employment and wage growth. Estimates are from eq. \ref{eq:reg_demp_daw}, with rolling regression coefficients and 95\% confidence intervals of 4-year windows, starting in 2003-2007. The dependent variables are annual growth rates of employment and wages. Employment regression includes the log of the initial period of employment. Wage regression includes a log of initial period employment, log initial period wage and log initial period wage squared. Occupation and sector fixed effects included.}}
    \label{fig:coefplot_demp_daw} 
\end{figure}

Figures \ref{fig:dist_te_emp_soc_sec} and \ref{fig:dist_te_emp_s} show the results. Overall, in 2023, 34\% of US employment is highly exposed to AI technologies, while medium and low exposure represent 32\% and 34\%, respectively.
Our findings do not suggest a polarizing effect of AI exposure as found by~\citeauthor{Frey-Osborne-17}; on the contrary, AI seems to have a more balanced impact on the labor market. This is because our indicator can capture recent advances in AI, such as LLMs, which have affected occupational groups such as management, business, administration, and finance, as well as ICT and science, which are intensive in non-routine cognitive tasks. For example, AI technologies are increasingly being used to diagnose diseases, write reports, code, or brainstorm ideas in management and business. On the contrary, previous studies that focus more on the impact of AI on routine tasks find that these tasks and occupations are less exposed to AI.  Grouping occupations by skill intensity shows that in the group with high exposure to AI, 88\% of employment is in high-skill jobs; in the group with medium exposure, 53\% of employment is in medium-skill jobs, while 40\% is in high-skill jobs. In the group with the lowest exposure to AI, 67\% are medium-skill jobs and 25\% low-skill jobs. Overall, AI exposure disproportionately affects high-skill jobs, which are characterized by the competencies most heavily affected by AI technologies. Next, we analyze the relationship between AI exposure and workers' characteristics. Figure \ref{fig:margins_biscatter} shows that \teai exposure is higher for workers with a high level of education, particularly graduates and postgraduates. There is a slight increase in exposure with age, although the variation in exposure is limited after age 30. Men are more exposed than women at all ages. Finally, we assess the relationship between AI exposure, employment, and wages. To compute the medium-term effect of AI in a flexible way, allowing for changes during the estimation period, we compute the log change in employment and wages over a 4-year rolling window from 2003 to 2023.\footnote{The last 4Y variation is therefore 2019-2023.} Therefore, for each 4Y window, we run the following:
\begin{equation}
\Delta y_{i,j} = \alpha +\beta TEAI_i + \bm{Z}^\prime_{i,j} \bm{\gamma} +\delta_i + \eta_j + \epsilon_{i,j}
\label{eq:reg_demp_daw}
\end{equation}

where $\Delta y_{i,j}$ denotes the 4-year change in log employment and log wages in sector $j$ for occupation $i$, $\bm{Z}$ is a column vector of controls, $\delta_i$ denote occupation dummies, $\eta_j$ sector dummies and $\epsilon_{i,j}$ the error term. To control for possible endogeneity and omitted variable problems, we include as controls the initial level of employment, the initial level of wage, and wage squared. We also include detailed sector (NAICS) and occupation (SOC) fixed effects in the regression and cluster the errors at the sector level. Figure \ref{fig:rolling_emp_wage} shows that exposure to AI positively correlates with employment and wage growth throughout the entire period. This suggests that AI technologies complement labor and increase productivity, thereby boosting employment and wages in occupations with greater exposure to AI. The presence of detailed controls at the industry and occupation level allows us to control for factors on the production side (changes in output across industries), on the demand side (changes in product demand across industries) and on the labor supply side (changes in employment across industries and occupations) that are unrelated to AI technologies and that could affect wages and employment. Moreover, the focus on a relatively short period isolates our results from long-term trends within industries and occupations. Therefore, the positive relationship between employment and wages and AI exposure should mean that occupations more exposed to AI have stronger employment and wage growth within the occupation and sector. Our results contrast with those obtained by~\citeauthor{Acemoglu-Autor_Hazell-Restrepo-22}, 
 and~\citeauthor{Webb-23}, who find a negative relationship between employment and wages. The potential reconciliation between our findings and theirs lies, on the one hand, in our construction of a different measure of AI exposure that emphasises more recent advances in AI. On the other hand, our analysis focuses on changes over the last 20 years, whereas theirs focuses on changes over several decades.

\begin{figure*}[h!]
    \centering
    \includegraphics[width=\linewidth]{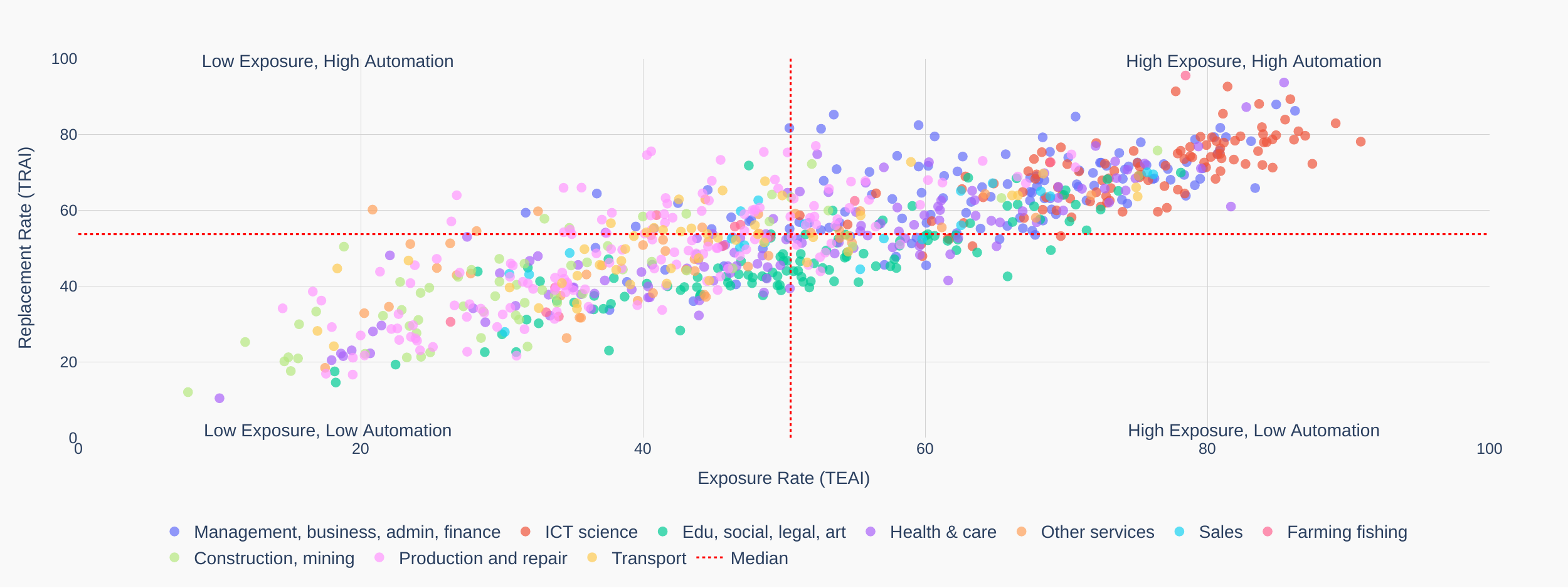}
  \caption{Task exposure and task substitutability. Each dot represents an O*NET occupation. Division lines are constructed using median values of the distribution of both indexes.  An \textbf{interactive demo} is available at \url{https://terminatoreconomy.com/home}}
  \label{fig:teai_replacement}
    
\end{figure*}

\subsection{Task Exposure or Substitutability?}

Whether AI will be an opportunity or a threat in the future will hinge on whether it complements or substitutes human labor. The measure we have developed so far is relatively agnostic in this respect, as we cannot yet disentangle substitutability from complementarity. In other words, a high \teai exposure measure does not necessarily imply full substitution of labor by technology, which, on the contrary, may rather complement human activities, leading to higher productivity without displacing labor. As described in section \ref{sec:ai-index} the \trai index is specifically designed to assess task replacement. 

\paragraph{What is the relationship between \trai and \teai index?} On the one hand, the \trai index can be considered a sub-product of the \teai index, as it is constructed using the motivation summaries that LLMs produce in developing the \teaiv. On the other hand, it differs because the task description that it analyzes is different (the \teai uses O$^*$NET task description \trai uses the summary of LLMs descriptions and motivations) and is designed to assess task substitutability, as the prompt asks LLM to specifically assess the possibility of replacing the human in task performance.
 
 Figure~\ref{fig:teai_replacement} displays the two indexes. Both scores have been rescaled on a scale of 0-100 for better readability.
 Please note that the scaling does not imply an exact percentage. For example, a score of 80 does not mean that there is an 80\% exposure or a replacement rate of 80\%.
 Also, while the figure shows different jobs, our replacement and exposure measures are constructed at \textit{task level}. Thus, even if an occupation has a high level of replacement, it does not mean that AI can completely replace it. Instead, it suggests that AI can perform a large number of tasks in that occupation, either completely or partially.

 The figure shows that the two measures are highly correlated, i.e., high values of the \teai index are generally associated with high values of the replacement index.
Grouping occupations improves the explanatory power of the analysis. For example, most ICT and science occupations display a high level of exposure and task substitutability. This means that for these occupations, AI can execute several tasks with a high degree of confidence. Human activity will still be required, and it is likely that these professionals will concentrate on more critical and high-value-added tasks (such as supervision, analysis, developing solutions, etc.), leaving other tasks to AI. 
Analogously, most teachers are characterized by a high level of exposure by a low level of substitutability: they are likely to work with AI, while AI will replace few tasks. Similarly, several health and care occupations are characterized by low levels of exposure and substitutability with few exceptions, such as ``Medical record specialists'' whose tasks are highly replaceable by technology.

 This provides some insights  about the complementarity vs substitutability debate. Complementarity is often interpreted as AI and humans sharing the same activities. When looking at specific tasks it emerges clearly that most of complementarity operates through substitution; in other words some tasks will be executed by AI and machines while humans will concentrate on others. AI and humans will complement each other in the same occupation, while task allocation within occupations is likely to change.

\section{Conclusions and Limitations\label{sec:conclusions}}
We proposed a methodology for assessing AI exposure using LLMs, with a fine-grained approach that analyses exposure for 923 jobs and their 19281 tasks, using the O*NET US official taxonomy as a baseline. Our approach differs from the literature in that we provide a data-driven evaluation, where LLM systems are asked to assess the suitability of tasks for AI, resulting in an occupation-based score of AI exposure and replacement. Our index of AI exposure, the so-called \teai is positively correlated with cognitive, problem-solving and management skills, highlighting the role of recent advances in AI that have a strong impact on management and decision-making tasks; conversely, our measure is negatively correlated with social skills, a known weakness of AI.

Regarding labor market outcomes, we find that AI exposure positively correlates with employment and wage growth over the period 2003-2023, suggesting that AI has a positive effect on productivity. Therefore, at least in the medium term, AI positively impacts the labor market. However, our estimates show that about one-third of the American workforce is at high risk of AI exposure, most in high-skilled jobs. For these workers, the overall impact of AI depends on whether exposure is associated with task substitutability.  We then construct a specific version of the AI index to assess task substitutability by AI, namely \traiv. We find that even in high-skill occupations, there is high variability in task substitution by AI, suggesting that AI and humans complement each other in the same occupation while allocating tasks within professions is likely to change.

A human evaluation has been performed to assess the approach's effectiveness, revealing that an average of 71-75\% of time evaluators agree with the score and explanations provided by our indexes, whilst only 2\% of time tend to disagree fully. To ensure full reproducibility, our methodology relies only on top-performing open LLMs, and all occupation and task lists are accessible through an online demo. This would allow the community to track future assessments of improvements in new LLM models while integrating our findings to help individuals and organizations shape policies for AI integration into careers and the workforce.

Finally, while O*NET offers a comprehensive and widely recognized taxonomy of the labor market, its static nature may not capture emerging tasks as they emerge in the real-labor market. To address this limitation, our approach is designed to be flexible and adaptable to any user-defined taxonomy, enabling organizations to assess the impact of AI within their operational contexts accurately.

\appendix
\section{Appendix}
\label{sec:appendix}
\subsection{Evaluating Consensus Among LLMs}
\label{consensus chap}
To construct a single indicator, we took a conservative approach by assigning to each task the value of the rating with the highest frequency among the three models; if the three ratings differed, we chose the lowest. To assess the agreement between the three rates expressed by the LLMs, for each single task, we compute a consensus metric \cite{TASTLE2007531}.
\begin{equation}
Cns(a_i) = 1 + \sum_{k=1}^{m} p_k \log_2 (1 - \frac{ \lvert LV_k - \mu_{LV} \rvert }{d_{LV}}) 
\label{eq:consensus}
\end{equation} 

 This process allows the three rates to be combined into a single measure of agreement. A higher consensus value indicates greater agreement between the observed rating values expressed for each task by the three different LLMs.
 The equation \eqref{eq:consensus} shows the consensus calculation in which $LV_k$ represents the observed rating value, $p_k$ its relative frequency, $\mu_{LV}$ represents the weighted average of the $LV$ ratings using $p_k$ probabilities as weights, and $d_LV$ represents the scale size of the ratings adopted. The logarithmic function computes the impact of the normalised difference between each rating and the weighted average, moderated by the $d_LV$ dimension. The calculation uses a repeated summation for each k-th rate expressed for each individual task. 
Analogously, to estimate the similarity between the motivations provided by the LLMs, we compute the centroid of semantic cosine similarity \cite{rahutomo2012semantic}, between them.
The embedding vectors for the centroid computation is obtained using an open source Transformer model: as for the LLMs, we chose the Transformer model to be used in accordance with the Massive Text Embedding Benchmark (MTEB) Leaderboard by~\cite{muennighoff-etal-2023-mteb}.\footnote{\url{https://huggingface.co/spaces/mteb/leaderboard}. Having English-language motivations, the choice fell on the \textit{UAE-Large-V1} \url{https://huggingface.co/WhereIsAI/UAE-Large-V1}, which represented an excellent compromise between effectiveness and efficiency, given its small size.}

\begin{figure}
    \centering
    \includegraphics[width=0.7\linewidth]{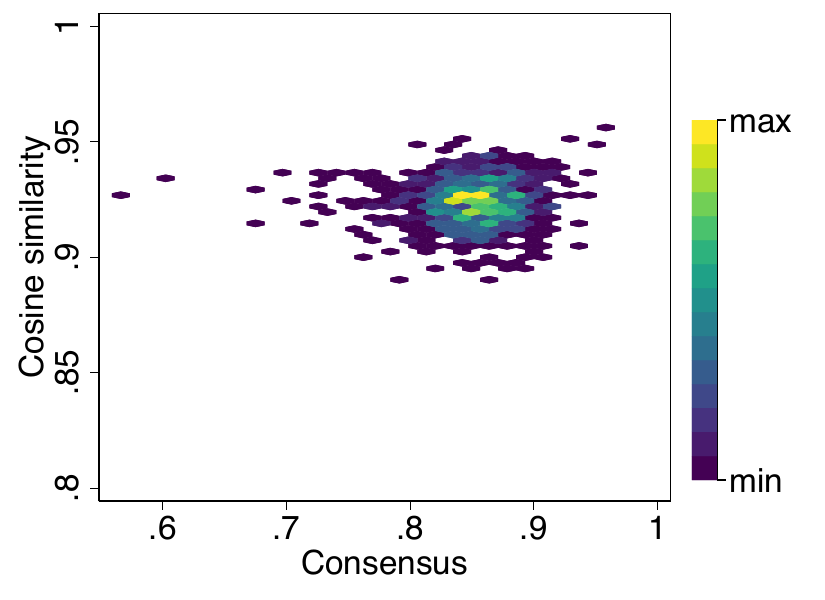}
    \caption{Heat map between cosine similarity of textual motivation of LLMs and consensus measure between scores. Data are aggregated at occupation level}
    \label{fig:heatmap}
\end{figure}

In particular, since higher cosine similarity values reflect higher semantic similarities between the text of the LLM motivations, we expect a strong correlation between the consensus metric and cosine similarity.
The heat map represented in Figure~\ref{fig:heatmap} shows that both the values of cosine similarity and the consensus metric are extremely high, with an average close to 0.9 in both cases.
On the one hand, this suggests coherence between LLM-generated rates and the associated motivations, and on the other hand it adds robustness to our 
conservative approach in choosing between different models.
\section*{Acknowledgements}
This paper is partially supported by the research activity of an EU H2020 Project \textit{PILLARS — Pathways to Inclusive Labor Markets} - Grant agreement 101004703 - \url{https://www.h2020-pillars.eu/}
\section*{Contribution Statement}
E.C., F.M., and M.M conceived the study; F.M., M.M., and A.S. developed the approach from a computational perspective E.C. performed the analyses from an economic perspective; F.M. managed the acquisition of funds. All authors wrote the initial manuscript draft, reviewed, and approved the final version of the manuscript. 
\bibliographystyle{named}
\bibliography{ijcai25}

\end{document}